\documentclass[a4paper,UKenglish,cleveref, autoref, thm-restate]{lipics-v2021}

\pdfoutput=1 
\hideLIPIcs  
\nolinenumbers

\title{Unifying Semantic Path Order and Weighted Path Order}

\author{Teppei Saito}
{JAIST, Japan}
{saito@jaist.ac.jp}
{https://orcid.org/0009-0001-9786-0044}
{JSPS KAKENHI Grant Number JP25KJ1363}

\author{Nao Hirokawa}
{JAIST, Japan}
{hirokawa@jaist.ac.jp}
{https://orcid.org/0000-0002-8499-0501}
{}

\authorrunning{T. Saito and N. Hirokawa}
\Copyright{Teppei Saito and Nao Hirokawa}


\category{} 

\relatedversion{} 

\acknowledgements{We thank Ren\'{e} Thiemann and Akihisa Yamada for the
valuable feedbacks on our work.  We also thank J\"{u}rgen Giesl for his
valuable questions at WST 2023 and FroCoS 2023.}

\EventEditors{John Q. Open and Joan R. Access}
\EventNoEds{2}
\EventLongTitle{42nd Conference on Very Important Topics (CVIT 2016)}
\EventShortTitle{CVIT 2016}
\EventAcronym{CVIT}
\EventYear{2016}
\EventDate{December 24--27, 2016}
\EventLocation{Little Whinging, United Kingdom}
\EventLogo{}
\SeriesVolume{42}
\ArticleNo{23}

\usepackage{arydshln}
\usepackage{url}
\usepackage{amssymb}
\usepackage{amsmath}
\usepackage{ebproof}
\usepackage{stmaryrd}
\usepackage{enumerate}
\usepackage{xspace}

\usepackage{pgfplots}

\usepackage{booktabs} 
\usepackage{tikz}
\usetikzlibrary{decorations,arrows}
\tikzstyle{E}=[dash pattern=on 2pt off 2pt]

\usepackage{supertabular} 

\newcommand{\NN}{\mathbb{N}}
\renewcommand{\AA}{\mathcal{A}}
\newcommand{\BB}{\mathcal{B}}
\newcommand{\RR}{\mathcal{R}}

\newcommand{\TT}{\mathcal{T}}
\newcommand{\VV}{\mathcal{V}}

\newcommand{\m}[1]{\mathsf{#1}}

\newcommand{\seq}[2][n]{{#2_1},\dots,{#2_{#1}}}

\newcommand{\lex}{\mathsf{lex}}

\newcommand{\spo}{\mathsf{S}}

\newcommand{\mspo}{\mathsf{MS}}

\newcommand{\wpo}{\mathsf{W}}

\newcommand{\gwpo}{\mathsf{G}}

\newcommand{\mgwpo}{\mathsf{MG}}

\newcommand{\DP}{\mathsf{DP}}

\newcommand{\emptylist}{[\,]}

\renewcommand{\emptyset}{\varnothing}

\newcommand{\AProVE}{\textsf{AProVE}\xspace}
\newcommand{\NaTT}{\textsf{NaTT}\xspace}

\newcommand\TTTT{%
 \textsf{T\kern-0.15em\raisebox{-0.55ex}T\kern-0.15emT\kern-0.15em\raisebox{-0.55ex}2}%
}
\newcommand\isafor{\textsf{Isa\kern-0.15exF\kern-0.15exo\kern-0.15exR}\xspace}
\newcommand\ceta{\textsf{C\kern-0.15exe\kern-0.45exT\kern-0.45exA}\xspace}

\begin{document}

\maketitle

\begin{abstract}
Monotonic semantic path orders and weighted path orders are powerful
reduction orders for proving termination of term rewrite systems.  
In this paper we present their simple unification as reduction orders and
reduction pairs.  We also discuss the use of it as ground total reduction
orders.
\end{abstract}

\section{Introduction}

We present a generalization of monotonic semantic path orders
(MSPOs)~\cite{BFR00} and weighted path orders (WPOs)~\cite{YKS15}.  In the
FroCoS 2023 paper~\cite{SH23} we showed that WPOs as reduction orders are
instances of MSPOs.  In this note we adopt the opposite approach:  First we 
introduce a generalization of WPOs dubbed GWPOs, and then show that MSPOs
are instances of GWPOs. We see that desirable properties of WPO carry over to GWPOs.  Full details
on these contributions are found in the first author's
dissertation~\cite{Sai26}.

\section{Reduction Orders}
\label{sec:ro}

This section considers \emph{reduction orders}, which
are well-founded orders on terms that are closed under contexts and substitutions.
It is known that a TRS $\RR$ is terminating if and only if there exists a
reduction order $>$ with $\RR \subseteq {>}$.

The basic ingredient of our unified orders (GWPOs) is
\emph{reduction triples}, originally introduced for MSPOs.
Formally, a triple $({\geqslant},{\sqsupseteq}, {\sqsupset})$ is a reduction triple if
$\geqslant$ is a rewrite preorder on terms,
$({\sqsupseteq},{\sqsupset})$ is a stable order pair on terms with $\sqsupset$ well-founded, and
$\sqsupseteq$ is harmonious to $\geqslant$.
Here, rewrite preorder means preorder (or quasi-order) closed under substitutions and contexts; 
the stability condition means that $\sqsupseteq$ and $\sqsupset$ are both closed under substitutions; and
the harmony condition means
that $t_i \geqslant u$ implies
$f(t_1,\ldots,t_i,\ldots,t_n) \sqsupseteq f(t_1,\ldots,u,\ldots,t_n)$
for all $n$-ary function symbols $f$, terms $t_1, \ldots, t_n, u$ and integers $1 \leqslant i \leqslant n$.

For example, given a reduction pair $({\geqslant}, {>})$ (which is a key
notion in the dependency pair method~\cite{AG00}),
the triple $({\geqslant}, {\geqslant}, {>})$ is a reduction triple.
Also, given a weakly monotone and well-founded algebra $\AA$ (say, a linear polynomial interpretation on natural numbers),
the triple $({\geqslant_\AA}, {\geqslant_\AA}, {>_\AA})$ is a reduction triple.
Moreover, given a precedence $\succeq$, the triple $({\TT \times \TT}, {\sqsupseteq}, {\sqsupset})$
is a reduction triple. Here, $\TT \times \TT$ is the universal relation relating every pair of terms,
and $\sqsupseteq$ and $\sqsupset$ are defined as comparison of head symbols w.r.t. $\succeq$.

We are ready to define the unified order GWPO.  Note that it generalizes GWPO in the sense of our earlier
paper~\cite{SH23}.

\begin{definition}
\label{def:gwpo}
Let $({\sqsupseteq_A}, {\sqsupset_A})$ and $({\sqsupseteq_B},
{\sqsupset_B})$ be pairs of relations on terms.  The \emph{generalized
weighted path order} $>_\gwpo$ (GWPO)
is defined as follows: 
$s >_\gwpo t$ if
\begin{enumerate}[1.]
  \item $s \sqsupset_A t$, or
  \item $s = f(\seq[m]{s}) \sqsupseteq_A t$ and one of the following conditions holds:
  \begin{enumerate}[a.]
    \item $s_i = t$ or $s_i >_\gwpo t$ for some 
    $1 \leqslant i \leqslant m$.
    \item $t = g(\seq[n]{t})$, $s >_\gwpo t_j$ for all $1 \leqslant j \leqslant n$,
    and moreover
    \begin{enumerate}[i.]
      \item $s \sqsupset_B t$, or
      \item $s \sqsupseteq_B t$ and $(\seq[m]{s}) >_\gwpo^\lex (\seq[n]{t})$.
    \end{enumerate}
  \end{enumerate}
\end{enumerate}
Now, let $({\geqslant_A}, {\sqsupseteq_A}, {\sqsupset_A})$ and $({\geqslant_B}, {\sqsupseteq_B}, {\sqsupset_B})$ be triples of relations on terms.
The \emph{monotonic generalized weighted path order} (MGWPO) $>_\mgwpo$ is the intersection
${\geqslant_A} \cap {\geqslant_B} \cap {>_\gwpo}$ where $>_\gwpo$ is induced by  $({\sqsupseteq_A}, {\sqsupset_A})$ and $({\sqsupseteq_B}, {\sqsupset_B})$.
\end{definition}

Technically, we take the intersection  ${\geqslant_A} \cap {\geqslant_B} \cap {>_\gwpo}$
because $>_\gwpo$ may not be closed under contexts. This technique is due to Borralleras, Ferreira and Rubio~\cite{BFR00}. 

To simplify the presentation, we globally assume that the signature is
finite. 
A relation $\leadsto$ on terms is said to have the \emph{subterm property}
if $s \leadsto t$ holds whenever $t$ is a proper subterm of $s$.

\begin{theorem}
\label{thm:mgwpo}
For reduction triples $({\geqslant_A}, {\sqsupseteq_A}, {\sqsupset_A})$ and $({\geqslant_B}, {\sqsupseteq_B}, {\sqsupset_B})$
with $\sqsupseteq_A$ having the subterm property,
the induced MGWPO $>_\mgwpo$ is a reduction order.
\end{theorem}
\begin{example}
We show the termination of the TRS 
\(
\RR = 
\{
\m{p}(\m{s}(x)) \to x,
\m{f}(\m{s}(x)) \to \m{f}(\m{p}(\m{s}(x)))
\}
\).
Let $\AA$ and $\BB$ be the algebras on
$\NN$ with 
$\m{p}_\AA(x) = \m{f}_\AA(x) = x$,
$\m{s}_\AA(x) = \m{f}_\BB(x) = \m{s}_\BB(x) = x + 1$, and
$\m{p}_\BB(x) = \max \{ 0, x - 1 \}$.
Then $({\geqslant_\AA},{\geqslant_\AA},{>_\AA})$ and
$({\geqslant_\BB},{\geqslant_\BB},{>_\BB})$ are reduction triples and
${\geqslant_\AA}$ has the subterm property.
We conclude that $\RR$ is terminating because 
the induced MGWPO satisfies $\RR \subseteq {>_\mgwpo}$. 
\end{example}

A WPO is parameterized by an algebra $\AA$ (assumed weakly simple), and a precedence $\succeq$.
The WPO is identical to the MGWPO whose reduction triple $A$ is induced from $\AA$
and $B$ is induced by $\succeq$.
(So actually, the subterm property of $\sqsupseteq_A$ assumed in \cref{thm:mgwpo} is an analog of
the weak simplicity of $\AA$ required by the WPO.)

The same goes for the MSPO:
An MSPO is parameterized by a \emph{single} reduction triple $({\geqslant},{\sqsupseteq}, {\sqsupset})$.
The MSPO is identical to the MGWPO 
whose reduction triple $A$ is the trivial one $({\TT \times \TT}, {\TT \times \TT}, {\emptyset})$
and $B$ is $({\geqslant},{\sqsupseteq}, {\sqsupset})$. Equivalently, the MSPO can be defined as follows.

\begin{definition}
\label{def:spo}
Let $({\sqsupseteq}, {\sqsupset})$ be a pair of relations on terms.
The \emph{semantic path order} $>_\spo$ is defined on terms as follows:
$s >_\spo t$ if $s = f(\seq[m]{s})$ and one of the next conditions holds.
\begin{enumerate}[a.]
    \item
    $s_i = t$ or $s_i >_\spo t$ for some $1 \leqslant i \leqslant m$. 
    \item
    $t = g(\seq{t})$ and $s >_\spo t_j$ for all $1 \leqslant j \leqslant n$,
    and moreover
    \begin{enumerate}[i.]
    \item
    $s \sqsupset t$, or
    \item
    $s \sqsupseteq t$ and $(\seq[m]{s}) >_\spo^{\lex} (\seq{t})$.
    \end{enumerate}
\end{enumerate}
For a reduction triple $({\geqslant},{\sqsupseteq}, {\sqsupset})$,
The \emph{monotonic semantic path order} $s >_\mspo t$ is defined as the intersection ${\geqslant} \cap {>_\spo}$,
where $>_\spo$ is induced by $({\sqsupseteq}, {\sqsupset})$.
\end{definition}

\begin{theorem}[\cite{BFR00}]
\label{thm:mspo}
Every MSPO is a reduction order.
\end{theorem}

Conversely, MGWPOs can be seen as instances of MSPOs.

\begin{theorem}
\label{thm:mgwpo-mspo}
Let $>_\mgwpo$ be the MGWPO
induced by reduction triples $({\geqslant_A}, {\sqsupseteq_A}, {\sqsupset_A})$ and $({\geqslant_B}, {\sqsupseteq_B}, {\sqsupset_B})$
with the subterm property of $\sqsupseteq_A$.
Then $>_\mgwpo$ and the MSPO $>_\mspo$ are identical,
where $>_\mspo$ is induced by $({\geqslant_A \cap \geqslant_B}, {\sqsupseteq_{AB}}, {\sqsupset_{AB}})$
and $({\sqsupseteq_{AB}}, {\sqsupset_{AB}})$ is the lexicographic combination of 
$({\sqsupseteq_{A}}, {\sqsupset_{A}})$ and $({\sqsupseteq_{B}}, {\sqsupset_{B}})$.
\end{theorem}

This is an analog of a main result of the FroCoS paper stating that WPOs are MSPOs~\cite{SH23}.
In fact, this result can be traced back to Geser's observation (\cite[Theorem~5]{Ges92})
that a version of the Knuth--Bendix order can be simulated by 
the semantic path order (SPO)~\cite{KL80}.

What is the point of having the unified order?
One way of defending our GWPO concerns \emph{ground totality}, which is a
crucial property for theorem proving methods like the superposition calculus.
On one hand, every WPO $>_\wpo$ satisfying totality of the precedence and the ordered algebra is ground total,
meaning that either $s >_\wpo t$ or $t >_\wpo s$ holds for all terms $s \neq t$.
On the other hand, this does not go naturally for the MSPO.
Based on this observation, we can construct 
a distinctive ground-total MGWPO.
For instance, we can construct a ground-total MGWPO $>_\mgwpo$
orienting the equations of the combinators $\m{S}$ and $\m{K}$:
\begin{align*}
(x \cdot z) \cdot (y \cdot z)
& {}>_\mgwpo  ((\m{S} \cdot x) \cdot y) \cdot z
&
(\m{K} \cdot x) \cdot y & >_\mgwpo x
\end{align*}
Moreover, such an MGWPO can be found by employing the standard techniques for automating path orders.
To our knowledge, this is not possible by existing constructions.
See \cite[Section~3.3]{Sai26} for the details of the construction.

Another point, again crucial for theorem proving, is that
the MGWPO provides a way to optimize comparison by an MSPO:
Given an instance $>_\gwpo$ of the GWPO,
one can conclude $s >_\gwpo t$ immediately from a strict decrease $s \sqsupset_A t$,
without recursive checks.
This is not possible for the simulating SPO $>_\spo$, compare \cref{def:gwpo,def:spo}.
Although this does not improve worst-case complexity, 
one can actually improve it by further imposing the subterm property 
on the \emph{strict} order $\sqsupset_A$.

\begin{proposition}
\label{prop:kbo-like-gwpo}
Let $({\sqsupseteq_A}, {\sqsupset_A})$ and $({\sqsupseteq_B}, {\sqsupset_B})$ be order pairs 
and $>_\gwpo$ the GWPO induced by them.
Suppose that $\sqsupset_A$ and $\sqsupseteq_A$ have the subterm property.
Then $s >_\gwpo t$ if and only if
\begin{enumerate}[1.]
  \item $s \sqsupset_A t$, or
  \item $s = f(\seq[m]{s}) \sqsupseteq_A g(\seq[n]{t}) = t$,  and one of the following conditions holds:
    \begin{enumerate}[i.]
      \item $s \sqsupset_B t$, or
      \item $s \sqsupseteq_B t$ and $(\seq[m]{s}) >_\gwpo^\lex (\seq[n]{t})$
    \end{enumerate}
\end{enumerate}
\end{proposition}

\cref{prop:kbo-like-gwpo} provides a definition of the GWPO 
in a similar fashion to the Knuth--Bendix order (KBO).
The KBO runs in linear time
while the lexicographic path order (LPO) runs in quadratic time (even with the best known algorithm).
Combining these, we obtain a procedure for deciding $s >_\mgwpo t$ in
\emph{linear} complexity in $|s| + |t|$,
where the measure is the number of comparisons by the base relations of the reduction triples $A$ and $B$.
This is not immediate from the definition of the MSPO which is similar to the LPO.

\section{Reduction Pairs}
\label{sec:rp}

A \emph{reduction pair} $({\geqslant}, {>})$ is an order pair defined on terms such that 
$\geqslant$ is a rewrite preorder and $>$ is a well-founded order closed under substitutions.
In its simplest form, the dependency pair method
states that a TRS $\RR$ is terminating
if and only if there is a reduction pair $({\geqslant}, {>})$
such that $\RR \subseteq {\geqslant}$ and $\DP(\RR) \subseteq {>}$,
where $\DP(\RR)$ is the set of dependency pairs, see~\cite{AG00,Der04}.

Let us consider a reduction pair version of the MGWPO incorporating 
\emph{partial statuses}, as is done for the original WPO~\cite{YKS15}. 
A partial status $\pi$ is a function that maps each $n$-ary function symbol to
a list $[i_1, \ldots, i_m]$ with $1 \leqslant i_1 < \cdots < i_m \leqslant n$. 
For an $n$-ary function symbol $f$ and terms $t_1, \ldots, t_n$,
we write $\pi(f)(t_1, \ldots, t_n)$ for the list $[t_{i_1}, \ldots, t_{i_m}]$
where $\pi(f) = [i_1, \ldots, i_m]$, and 
$\pi$ is called \emph{total} if $\pi(f) = [1,\ldots, n]$
for all $n$-ary function symbols $f$. 

\begin{definition}
\label{def:gwpo-reduction-pair}
Let $\pi$ be a partial status, and
let $({\geqslant_A}, {\sqsupseteq_A}, {\sqsupset_A})$ and $({\geqslant_B}, {\sqsupseteq_B}, {\sqsupset_B})$ be triples of relations on terms.
The \emph{generalized weighted path order}
$({\geqslant_\gwpo}, {>_\gwpo})$ is a pair of relations on terms defined simultaneously as follows:
$s \geqslant_\gwpo t$ if 
\begin{enumerate}[1.]
\item $s \sqsupset_A t$ or
\item $s \sqsupseteq_A t $ and one of the following conditions holds.
\begin{enumerate}[a.]
\item
$s = f(\seq[m]{s})$ and $s_i \geqslant_\gwpo t$ for some $i \in
\pi(f)$.
\item
$s = f(\seq[m]{s})$, $t = g(\seq[n]{t})$, $s >_\gwpo t_j$ for all 
$j \in \pi(g)$, and
\begin{enumerate}[i.]
\item
$s \sqsupset_B t$ or
\item
$s \sqsupseteq_B t$ and
$\pi(f)(\seq[m]{s}) \geqslant_\gwpo^\lex \pi(g)(\seq[n]{t})$.
\end{enumerate}
\item
$s \in \VV$ and $s = t$
\end{enumerate}
\end{enumerate}
The relation $>_\gwpo$ is defined by cases 1, 2a and 2b
where $\geqslant_\gwpo^\lex$ is replaced by $>_\gwpo^\lex$.
Here $({\geqslant_\gwpo^\lex}, {>_\gwpo^\lex})$ is the lexicographic extension of
$({\geqslant_\gwpo}, {>_\gwpo})$.
The \emph{monotonic generalized weighted path order}
$\geqslant_\mgwpo$
is defined as ${{\geqslant_A} \cap {\geqslant_B} \cap {\geqslant_\gwpo}}$,
where $({\geqslant_\gwpo}, {>_\gwpo})$ is induced by  $({\sqsupseteq_A}, {\sqsupset_A})$ and $({\sqsupseteq_B}, {\sqsupset_B})$.
\end{definition}

For a partial status $\pi$,
a relation $\leadsto$ on terms is \emph{simple with respect to} $\pi$
if $f(t_1, \ldots, t_n) \leadsto t_i$ for all $n$-ary function symbols $f$,
terms $\seq[n]{t}$ and $i \in \pi(f)$.
This is a generalized version of the subterm property corresponding
to the notion of $\pi$-simple algebra for WPOs.

\begin{theorem}
\label{thm:gwpo-reduction-pair}
Let $\pi$ be a partial status, and
$({\geqslant_A}, {\sqsupseteq_A}, {\sqsupset_A})$ and $({\geqslant_B}, {\sqsupseteq_B}, {\sqsupset_B})$ reduction triples.
If $\sqsupseteq_A$ is simple with respect to $\pi$,
then $({\geqslant_\mgwpo}, {>_\gwpo})$ is a reduction pair.
\end{theorem}

Again, WPOs as reduction pairs can be constructed as GWPOs
by constructing $A$ from algebras and $B$ from precedences.
While we are not aware of a dedicated version of (M)SPOs as reduction
pairs in any preceding work, it can be defined as GWPO whose first
reduction triple is trivial.

\begin{definition}
\label{def:spo-reduction-pair}
Let $\pi$ be a partial status, and
$({\geqslant}, {\sqsupseteq}, {\sqsupset})$ a reduction triple.
The \emph{semantic path order} $({\geqslant_\spo}, {>_\spo})$
induced by $\pi$ and $({\sqsupseteq}, {\sqsupset})$
is defined inductively as follows: $s \geqslant_\spo t$ if
\begin{enumerate}[1.]
\item
$s = f(\seq[m]{s})$ and $s_i \geqslant_\spo t$ for some $i \in
\pi(f)$.
\item
$s = f(\seq[m]{s})$, $t = g(\seq[n]{t})$, $s >_\spo t_j$ for all 
$j \in \pi(g)$, and
\begin{enumerate}[a.]
\item
$s \sqsupset t$ or
\item
$s \sqsupseteq t$ and
$\pi(f)(\seq[m]{s}) \geqslant_\spo^\lex \pi(g)(\seq[n]{t})$.
\end{enumerate}
\item
$s \in \VV$ and $s = t$
\end{enumerate}
The relation $>_\spo$ is defined by cases 1, 2a and 2b
where $\geqslant_\spo^\lex$ is replaced by $>_\spo^\lex$.
Here $({\geqslant_\spo^\lex}, {>_\spo^\lex})$ is the lexicographic extension of
$({\geqslant_\spo}, {>_\spo})$.
The \emph{monotonic semantic path order}
$\geqslant_\mspo$ is defined as ${\geqslant} \cap {\geqslant_\spo}$,
where $({\geqslant_\spo}, {>_\spo})$ is induced by  $({\sqsupseteq}, {\sqsupset})$.
\end{definition}

\begin{corollary}
\label{cor:mspo-reduction-pair}
Let $\pi$ be a partial status, and
$({\geqslant}, {\sqsupseteq}, {\sqsupset})$ a reduction triple.
Then $({\geqslant_\mspo}, {>_\spo})$ is a reduction pair.
\end{corollary}

For a term $t = f(\seq{t})$ we define $t^!$ as $f^!(\seq{t})$.  Here $f^!$
is a fresh function symbol.  Let $\AA$ be a weakly monotone well-founded
algebra.  We write $s \sqsupseteq_\AA t$ if either $s$ and $t$ are the same
variable or they are not variables and
$s^! \geqslant_\AA t^!$. Similarly, we write $s^! >_\AA t^!$ if $s$ and $t$ are not
variables and $s^! >_\AA t^!$.
Then $({\geqslant_\AA},{\sqsupseteq_\AA},{\sqsupset_\AA})$ is a reduction triple.

\begin{example}
\label{ex:spo}
Consider the TRS $\RR$ taken from the Termination Problem Database
(\texttt{TRS\_Standard/Zantema\_05/z08}):\footnote{\url{https://github.com/TermCOMP/TPDB-ARI/}} 
\begin{align*}
\m{f}(\m{a}, \m{f}(\m{b}, \m{f}(\m{a}, x))) & \to \m{f}(\m{a}, \m{f}(\m{b}, \m{f}(\m{b}, \m{f}(\m{a}, x))))
&
\m{f}(\m{b}, \m{f}(\m{b}, \m{f}(\m{b}, x))) & \to \m{f}(\m{b}, \m{f}(\m{b}, x))
\end{align*}
The set $\DP(\RR)$ consists of the two dependency pairs:
\begin{align*}
\m{F}(\m{a},\m{f}(\m{b}, \m{f}(\m{a}, x))) &\to \m{F}(\m{a}, \m{f}(\m{b}, \m{f}(\m{b}, \m{f}(\m{a}, x))))
&
\m{F}(\m{a}, \m{f}(\m{b}, \m{f}(\m{a}, x))) \to \m{F}(\m{b}, \m{f}(\m{b}, \m{f}(\m{a}, x)))
\end{align*}
Here, $\m{F}$ is the tuple symbol for $\m{f}$ introduced by the dependency pair method.
Let $\AA$ be the algebra on $\NN$ with
$\m{F}^{(!)}_\AA(x, y) = \m{f}_\AA(x,y) = x$,
$\m{f}^!_\AA(x,y) = y$,
$\m{a}^{(!)}_\AA = 1$, and
$\m{b}^{(!)}_\AA = 0$; 
and let $\pi$ be the partial status with $\pi(\m{F}) = [2]$ and
$\pi(\m{f}) = \pi(\m{a}) = \pi(\m{b}) = \emptylist$.  
The reduction pair $({\geqslant_\mspo},{>_\spo})$ induced from 
$({\geqslant_\AA},{\sqsupseteq_\AA},{\sqsupset_\AA})$ satisfies 
$\RR \subseteq {\geqslant_\mspo}$ and $\DP(\RR) \subseteq {>_\spo}$,
from which we conclude that $\RR$ is terminating.
In particular, $\m{F}(\m{a},\m{f}(\m{b}, \m{f}(\m{a}, x))) >_\spo \m{F}(\m{a}, \m{f}(\m{b}, \m{f}(\m{b}, \m{f}(\m{a}, x))))$
follows by case~2b and $\m{f}(\m{b}, \m{f}(\m{a}, x)) >_\spo \m{f}(\m{b}, \m{f}(\m{b}, \m{f}(\m{a}, x)))$,
which in turn follows from case~2a and $\m{f}^{!}(\m{b}, \m{f}(\m{a}, x)) >_\AA \m{f}^{!}(\m{b}, \m{f}(\m{b}, \m{f}(\m{a}, x)))$.
The last inequality is confirmed by calculation:
\[
\m{f}^{!}_\AA(\m{b}_\AA, \m{f}_\AA(\underline{\m{a}_\AA}, x)) = 1 > 0 =  \m{f}^{!}_\AA(\m{b}_\AA, \m{f}_\AA(\underline{\m{b}_\AA}, \m{f}_\AA(\m{a}_\AA, x)))
\]
The underlined parts correspond to the values $1$ and $0$.
Note that WPOs with linear polynomial interpretations are unable to satisfy those inclusions.
\end{example}

Next we investigate an analog of \cref{thm:mgwpo-mspo}.
Let $({\geqslant_\mgwpo}, {>_\gwpo})$ be the generalized weighted path order (\cref{def:gwpo-reduction-pair})
induced by a partial status $\pi$ and reduction triples $({\geqslant_A}, {\sqsupseteq_A}, {\sqsupset_A})$ and $({\geqslant_B}, {\sqsupseteq_B}, {\sqsupset_B})$, and 
let $({\geqslant_\mspo}, {>_\mspo})$ be the semantic path order (\cref{def:spo-reduction-pair})
induced by $\pi$ and $({\geqslant_A \cap \geqslant_B}, {\sqsupseteq_{AB}}, {\sqsupset_{AB}})$.

\begin{theorem}
\label{thm:mgwpo-mspo2}
Assume that $\sqsupseteq_A$ is simple with respect to $\pi$.
Then the inclusions
${\geqslant_\mspo} \subseteq {\geqslant_\mgwpo}$ and ${>_\spo} \subseteq {>_\gwpo}$ hold. 
If $\pi$ is total, then 
$({\geqslant_\mspo}, {>_\spo})$ is identical to $({\geqslant_\mgwpo}, {>_\gwpo})$.
\end{theorem}

For the second claim, the assumption that $\pi$ is total is essential.

\begin{example}
\label{ex:non-simulation}
Suppose that the signature only contains a unary function symbol $\m{f}$.
Let $\pi$ be the partial status with
$\pi(\m{f}) = \emptylist$, 
let $\AA$ be the algebra on $\NN$  defined by $\m{f}_\AA(x) = x + 1$,
and 
let $({\sqsupseteq_B}, {\sqsupset_B})$ be an arbitrary pair of relations
(e.g., one induced by a precedence to consider the original WPO).
On one hand, the GWPO $({\geqslant_\gwpo}, {>_\gwpo})$ induced by $({\geqslant_\AA}, {>_\AA})$
and $({\sqsupseteq_B}, {\sqsupset_B})$ satisfies $\m{f}(x) \geqslant_\gwpo x$ and $\m{f}(x) >_\gwpo x$ 
by case 1. On the other hand, no SPO $({\geqslant_\spo}, {>_\spo})$ (regardless of the underlying order pair)
satisfies neither $\m{f}(x) \geqslant_\spo x$ nor $\m{f}(x) >_\spo x$, because 
only possible case~1 is blocked by $\pi(\m{f}) = \emptylist$.
\end{example}

All in all, in the setting of \cref{thm:mgwpo-mspo2} in which $\sqsupseteq_A$ is simple with respect to $\pi$,
the GWPO is always preferred over the SPO.
In contrast, if the simplicity condition is not satisfied,
the SPO provides a way to construct a reduction pair (\cref{cor:mspo-reduction-pair}),
by combining the reduction triples $A$ and $B$ lexicographically into the single one.
In other words, the SPO allows us to virtually bypass the simplicity condition of \cref{thm:gwpo-reduction-pair}.

In order to evaluate the GWPO and SPO as reduction pairs, we implemented a
prototype termination tool based on the dependency pair method and the presented reduction pairs.
The problem set for the experiment consists of
1528 finite TRSs 
from version 11.5 of the Termination Problem Database,
and one-minute time limit is set for the benchmark.
Overall, one obtains more termination proofs by switching from the original WPO
to our methods, see~\cite[Section~4.5]{Sai26} for the details.
However, compared to state-of-the-art termination provers, 
every problem solved by our tool is also solved by \AProVE or \NaTT.
So the authors do not know if any new problem can be solved by our methods.
 
To grasp the evaluation results, we pick up
the following three kinds of reduction pairs:
\begin{itemize}
\item
\label{wpo}
the WPO induced by a linear polynomial interpretation and a
precedence;
\item
\label{gwpo}the GWPO (\cref{thm:gwpo-reduction-pair}) induced by reduction triples $A$
and $B$; and
\item\label{spo}
the SPO (\cref{cor:mspo-reduction-pair}) induced by the lex combination of
reduction triples $A$ and $B$.
\end{itemize}
Here, $A$ is induced by a linear polynomial interpretation  on natural numbers,
and for linear polynomial interpretations $\AA$ (including those for the WPO)
we demand $a_1, \ldots, a_n \in \{ 0, 1 \}$ for each interpretation $f_\AA(x_1, \ldots, x_n) = a_0 + a_1 x_1 + \cdots + a_n x_n$.
Only for the WPO and the GWPO, we additionally impose simplicity with respect to the partial status.
The other reduction triple $B$ is induced by a max/plus interpretation $\BB$ on natural numbers
in which $b_1, \ldots, b_n \in \{ 0, 1 \}$ and $c_0 \geqslant 0$ are required
for each interpretation $f_\BB(x_1, \ldots, x_n) = \max \{ c_0, b_1 x_1 + c_1, \ldots, b_n x_n + c_n \}$.
For $\BB$, the integers $c_1, \ldots, c_n$ are allowed to be negative.
So interpretations like $f_\BB(x) = \max \{0, x - 1 \}$ belong to this
class.  We construct $A$ and $B$ for the SPO and $B$ for the GWPO
in the way described before \cref{ex:spo}, using fresh symbols $f^!$.
Note that this construction cannot be used for $A$ of the GWPO due to the simplicity requirement.

In this setting, the WPO, the GWPO, and the SPO (each of which is
used with the dependency pair framework) prove termination of 486, 591, and 595
systems, respectively. Actually, the three configurations solve different sets of problems.
There are two reasons:
\begin{itemize}
\item
In theory, the GWPO subsumes the WPO as the max/plus interpretation
generalizes precedence.  However, due to the one-minute time limit
imposed in our experiment, the GWPO misses 7 problems handled by the
WPO.
\item
Already in theory, the SPO considered here is incomparable with the WPO and
the GWPO. In fact, the SPO does not require $\AA$ to be weakly simple
with respect to the partial status.  So \cref{thm:mgwpo-mspo2} does
not apply; see also \cref{ex:non-simulation}.
\end{itemize}

\bibliography{references}

\end{document}